# Comment on "Biological modeling of gold nanoparticle enhanced radiotherapy for proton therapy" by Lin et al. [Phys. Med. Biol. 60 (2015) 4149–4168]

Hans Rabus [1]

[1] Physikalisch-Technische Bundesanstalt (PTB), Berlin, Germany

E-mail: hans.rabus@ptb.de

**Abstract**

In their article published in Phys. Med. Biol. 60 (2015) 4149–4168, Lin *et al* studied the radiosensitizing effect of gold nanoparticles (GNPs) using radiation transport simulations and a biological model for the survival of irradiated cells. This comment points out several caveats to the methodlogy used by Lin et al. that may not be evident to readers and may contribute to confusion in the literature about the radiation effects of gold nanoparticles. The two main caveats are the high mass fraction of gold considered and a potential problem with the modified local effect model used to predict cell survival.

---

## 1. Gold concentration

In the paper of Lin *et al* (2015), the main studied nanoparticle size and concentration of GNPs are 50 nm and 1 µM, respectively. Assuming that the mass density of gold in the GNPs is that of bulk gold, namely $\rho_{Au}$ = 19.32 g/cm³, a 50 nm GNP contains

$$(50\times10^{-7} \text{ cm})^3 \times \pi/6 \times 19.32 \text{ g/cm}^3/(196.97 \text{ g/mol}) \times 6.022\times10^{23} \text{ mol}^{-1} = 3.81\times10^6 \tag{1}$$

gold atoms. Thus, a concentration of 1 µM GNPs corresponds to a concentration of gold atoms of about 3.8 mol/L. This implies a mass density of gold in solution of 750 g/L, which corresponds to a mass fraction of gold of about 43%!

When irradiated with a 50 kVp photon spectrum, most photons have energies in the range where the mass-energy absorption coefficients of gold and water differ by two orders of magnitude (Hubbell and Seltzer 2004). Therefore, a photon fluence that produces an absorbed dose of 1 Gy in water in the absence of the GNPs results in an average dose of about 40 Gy when the GNPs are present. So it is not a big surprise that negligibly small survival rates are predicted for the 50 kVp spectrum!

For these low-energy photons, Lin et al. (2015) also investigated the dependence on GNP concentration in the range between 10 nM and 1 µM. From the argument presented above, a GNP concentration of 10 nM corresponds to a mass fraction of gold of about 0.75%, which is still high but closer to the range of realistic values. For the linac spectrum and protons, on the other hand, the increase in average absorbed dose is much smaller. Here, Lin et al. (2015) studied concentrations between 100 nM and 10 µM, corresponding to mass densities of gold in solution between 75 g/L and 7.5 kg/L and mass fractions between 7% and 88%! These are definitely unrealistically high values.

## 2. Inconsistencies in the description of the simulation setup

Apart from the issue of high GNP concentration, the data in the "Materials and Methods" section of the paper appear contradictory. The paper states, "A concentration of 1 µM using 50 nm diameter GNPs results in $1.4 \times 10^5$ GNPs for the Nucleus, CellHomo and Cytoplasm geometries (based on a cylindrical volume of 13.5 µm diameter and 2 µm thickness)." The three geometries refer to the cases where the GNPs are located only in the cell nucleus, uniformly distributed throughout the cell, and only in the cytoplasm. It is obviously impossible for the same number of GNPs to correspond to the same concentration in all three cases. For a given concentration, the number of GNPs must be different for the cell and for the cell nucleus, simply because the cell has a larger volume.

A cylinder with a diameter of 13.5 µm and a height of 2 µm has a volume $V_c$ of

$$V_c = (13.5 \text{ µm})^2 \times \pi/4 \times 2 \text{ µm} = 2.86 \times 10^2 \text{ µm}^3 = 2.86 \times 10^{-13} \text{ L} \tag{2}$$

At a concentration $c_{GNP}$ of nanoparticles of 1 µM, the number $N_{GNP,c}$ of GNPs in the cell is given by

$$N_{GNP,c} = c_{GNP} \times V_c \times N_A = 1 \times 10^{-6} \text{ mol/L} \times 2.86 \times 10^{-13} \text{ L} \times 6.02 \times 10^{23} \text{ mol}^{-1} = 1.72 \times 10^5.$$

Conversely, if the number of GNPs in the nucleus, $N_{GNP,n}$, is $1.4 \times 10^5$ and $c_{GNP} = 1$ µM, then the volume $V_n$ of the nucleus is

$$V_n = N_{GNP,n} / (c_{GNP} \times N_A) = 1.4 \times 10^5 / (1 \times 10^{-6} \text{ mol/L} \times 6.022 \times 10^{23} \text{ mol}^{-1}) = 2.33 \times 10^{-13} \text{ L} = 233 \text{ µm}^3 \tag{3}$$

An 8 µm diameter circle has an area of $(8 \text{ µm})^2 \times \pi/4 = 50.3 \text{ µm}^2$, so a cylindrical cell nucleus of volume $V_n = 233$ µm³ has a height of 4.64 µm, which exceeds the cell's assumed thickness of 2 µm. If the nucleus is assumed to be spherical with a diameter of 8 µm, its volume $V_n$ is

$$V_n = (8 \text{ µm})^3 \times \pi/6 = 2.68 \times 10^2 \text{ µm}^3 = 2.68 \times 10^{-13} \text{ L} \tag{4}$$

and $N_{GNP,n} = 1.4 \times 10^5$ corresponds to a GNP concentration of

$$c_{GNP} = N_{GNP,n}/V_n/N_A = 1.4 \times 10^5 / (2.68 \times 10^{-13} \text{ L} \times 6.022 \times 10^{23} \text{ mol}^{-1}) = 0.87 \text{ µM}. \tag{5}$$

It should be noted that a sphere with a diameter of 8 µm will not fit into a cylinder 2 µm high, and that the volumes given in Eqs. 2 and 4 are similar but not identical. It therefore remains unclear what geometry and concentration of GNPs was actually used.

## 3. Local effect model

Section 2.3 of (Lin *et al* 2015) describes a variant of the local effect model (LEM), called GNP-LEM, which uses a dose distribution composed of the dose contribution from interactions in water and the localized additional dose contribution around GNPs. The paper states that the latter dose contribution is obtained "by multiplying the dose from a single ionizing event by the number of GNPs, the interaction probability per Gray and the prescribed dose" and that "The GNP-LEM developed in this study was implemented in 2D, where the volume integration is reduced to an area integration over the cell nucleus."

It is not clear what these two statements actually mean. The first statement suggests that the spatial arrangement of the GNPs was not taken into account. The second statement suggests that GNPs are treated in analogy to ion beams in the original LEM, where the dose distribution has a cylindrical symmetry around the ion trajectory. If one then performs the integral over a plane perpendicular to this trajectory, one obtains the number of lesions produced per pathlength of the ion. For ions with low energy loss in the nucleus and a nucleus with cylindrical shape irradiated along the cylinder axis, the total number of lesions is obtained by multiplying the cylinder height with the number of lesions produced per pathlength.

How this can be applied to GNPs is not clear. In this context, it should be mentioned that the formula given in the article of Lin *et al* (2015) for the total number of lethal lesions (second formula on page 4149) is incorrect because the logarithm of the survival probability (appearing in the first formula on page 4149) is missing. The correct formula is

$$\bar{N}_{total} = \int_V \frac{-\ln S_x(d(x,y,z))}{V} dV \tag{6}$$

Since the procedure used calculate the integral is not described in sufficient detail, it is not possible to assess whether or not "area integration over the cell nucleus" gives a correct evaluation of the total number of induced lesions. In conjunction with the first unclear statement, there is a possibility that Lin *et al* (2015) implicitly assumed (as did Jones *et al* (2010)) that a two-dimensional projection of the dose distributions around GNPs onto a plane and integration over that plane would provide them with the same information as a three-dimensional integral. However, as pointed out in (Rabus *et al* 2021), such an approach implies that it does not determine the dose enhancement, or the number of lesions produced by GNPs. Instead, such an approach determines these quantities in the case where the GNPs are replaced by cylindrical rods of gold, that have the same circular cross section as the GNPs but a length equal to the thickness of the nucleus. The resulting integration value greatly overestimates the number of lethal lesions and therefore leads to an underestimation of cell survival.

Whether the results of (Lin *et al* 2015) suffer from this deficiency cannot be judged, as their paper does not include detailed information on how they actually proceeded.

## 4.   Dependence of dose per ionization on GNP size

In Section 3.2 of (Lin *et al* 2015), the authors comment on the dependence of the dose contribution from electrons produced in ionizations in the GNP on the GNP size, which can be seen in their Fig. 4. Their explanation is, "For the same energy absorbed by a single GNP, the secondary electrons generated in a large GNP are more likely to lose their energy before reaching the surface. Such self-absorption contributes to the lower dose deposited around the GNP by one ionization event for larger GNPs."

The main reason for the difference in dose contribution between different GNP sizes is that the mass of a water shell of the same thickness around GNPs of different size increases with the square of the GNP radius. Therefore, one would expect the dose at the surface of a 2 nm GNP to be 625 times higher than at the surface of a 50 nm GNP. That the authors only find an increase by a factor 215 suggests that contrary to the authors' claim, the higher number of interactions in a larger GNP actually increases the dose contribution outside.

## Conclusions

Most of the results shown in (Lin *et al* 2015) are for gold concentrations that appear unrealistically high. The trend of decreasing survival probability with decreasing GNP size for the same amount of gold in the cells, shown in the left panel of Fig. 8 of (Lin *et al* 2015), should also apply for realistic gold concentrations. If the results shown in the right panel of Fig. 8 for 2 nm GNPs apply to a concentration of 1 µM of these GNPs, the corresponding concentration of gold atoms is 250 µM or 50 mg/L, which corresponds to a gold mass fraction of $5 \times 10^{-5}$. Therefore, the curve for 2 nm GNPs in the right panel of Fig. 8 presumably indicates a realistic magnitude of effects from GNPs during proton irradiation, if the authors' calculations are not compromised by the potential problem described in Section 3. It should be noted, however, that even if their calculations of cell survival are correct, the 2 nm GNP data shown in the right panel of Fig. 8 only apply to the case that survival is determined solely by physical dose enhancement and not by other factors, such as chemical and biological effects of GNPs.

## Supplement

This comment was submitted (without this supplement) to Physics in Medicine and Biology in December 2022, where it was reviewed by Editorial board members who recommended the comment to be rejected. Since the arguments raised by the reviewers suggest in my view that they considered the approach of writing this comment as disrespectful to the authors or even may have taken offense themselves my suspecting malevolent intentions on my side, I would like to add here the following statements for clarification:

- ***The intent of this comment is not to blame the authors of the article or the peer-review process for this article.***
- ***In my view, science is a collective learning effort and peer review is the best approach we have to assure quality in this process.***

I assume that the members of the editorial board were driven by a - very commendable - desire to protect their authors from disrespectful attacks, and probably had serious concerns about a mudslinging match might break out between the commentator and the authors. The former was not intended, and the latter is not imminent, as I pointed out in my appeal against the first rejection notice. I therefore personally feel that the arguments on which the first and the final decision were based were disrespectful to me. The interested reader may refer to the attached reproduction of the email correspondence.

---

**Your manuscript PMB-114487 - Decision on your manuscript**
Physics in Medicine and Biology
An hans.rabus
Kopie hans.rabus

06.01.2023 12:24

Details verbergen

| | |
|---|---|
| Von | "Physics in Medicine and Biology" <onbehalfof@manuscriptcentral.com> |
| An | hans.rabus@ptb.de |
| Kopie | hans.rabus@ptb.de |
| Protokoll | Bitte Antwort an pmb@ioppublishing.org<br>Diese Nachricht wurde beantwortet. |

Dear Dr Rabus,

Re: "Comment on "Biological modeling of gold nanoparticle enhanced radiotherapy for proton therapy" by Lin et al. [Phys. Med. Biol. 60 (2015) 4149–4168]"

Manuscript reference: PMB-114487

The Editorial Board of Physics in Medicine and Biology has considered your Comment. Unfortunately, they have recommended that we should not publish your work.

You can find the reasons for this decision in the report(s). These can be found below and/or attached to this message.

We are sorry that we cannot respond more positively but would like to thank you for your interest in Physics in Medicine and Biology.

REVIEWER REPORT(S):

Referee: 1

COMMENTS TO THE AUTHOR(S)
EDITORIAL BOARD PRELIMINARY REPORT

The submission of comments on a several-year old paper may not be timely enough. I would recommend the author to send a personal letter to the authors of the objected paper to give them a chance to handle the problems or to write his own paper to draw new valid conclusions.

Letter reference: DEC:RejBM:S

Dear Editor,

thank you for your email. It is understood that given this reviewer comment, you had no other choice than rejecting the manuscript.

However, may I point out that the reviewer's comment suggests that they have not read the manuscript? The recommendation appears to be based solely on the fact that the paper commented on was published several years ago and on the assumption that I did not contact the authors before writing the comment.

The former is a fact, the paper appeared years ago. It has since then inspired other authors to follow suit and run into the same pitfalls as Lin et al did, as I wrote in my cover letter.

Writing a comment now - that finally someone noticed the issues - seemed timely from my perspective and the best way to post a warning sign directly linked to the paper that set the bad model. (There is no offense meant with this judging statement.) In science, it should be never too late for pointing out errors in results or caveats of methodologies.

The reviewer's advice to write an own paper also comes years too late. I wrote several papers (co-authored by one of the authors of the paper I tried to comment on) on methodological pitfalls with nanoparticle simulations. Which are not stopping people from using papers like the one of Lin et al. as a starting point for their investigations, unaware of the caveats.

Finally, contrary to the reviewer's assumption, I did contact the two co-authors of the paper, which I know in person, and we considered writing the Comment together. This appeared not to fit the intention of Comments in PMB, as appeared strange if they coauthored the Comment and then wrote a reply.

I write this long email to you for two reasons.

1. One is to ask advice how to handle this particular case. I assume there is no way of appeal. Is it possible for co-authors of the paper commented on to co-author the comment on this paper?

In this context I should inform you that the preprint was uploaded to arXiv, simply to claim priority over the authors of the manuscript I recently reviewed, which prompted me to write my comment. As usual, I delivered a very detailed reviewer report outlining all the problems they have inherited from Lin et al. by copying their methodology. Nevertheless, I think it would be more appropriate for the comment to appear in PMB.

2. The second reason is to inquire whether the PMB Editorial Board has a position on how to deal with cases where methodological problems with papers are discovered long after publication?

PMB has a rigorous peer review process, which is the best we can do to assure scientific quality. But it cannot be expected to be perfect. As a physicist with a long track record in experiments before my career as simulator, I repeatedly made the experience that other reviewers of simulation studies tended to be satisfied when the code details are comprehensively described, but often failed to see obvious implausibilities in the methodology.

And every published paper is a part of the scientific literature that can influence future researchers in one way or the other. So keeping the piece is not an option in such cases, in my view.

Thank you for your attention. I am looking forward to your answer.

With kind regards

Hans

**Your manuscript PMB-114487 - Decision on your manuscript**
Physics in Medicine and Biology

Dear Dr Rabus,

Re: "Comment on "Biological modeling of gold nanoparticle enhanced radiotherapy for proton therapy" by Lin et al. [Phys. Med. Biol. 60 (2015) 4149–4168]"

Manuscript reference: PMB-114487

The Editorial Board of Physics in Medicine and Biology has considered your Comment. Unfortunately, they have recommended that we should not publish your work.

You can find the reasons for this decision in the report(s). These can be found below and/or attached to this message.

We are sorry that we cannot respond more positively but would like to thank you for your interest in Physics in Medicine and Biology.

REVIEWER REPORT(S):

Referee: 1

COMMENTS TO THE AUTHOR(S)
The submission of comments on a several-year old paper may not be timely enough. I would recommend the author to send a personal letter to the authors of the objected paper to give them a chance to handle the problems or to write his own paper to draw new valid conclusions.

Referee: 2

COMMENTS TO THE AUTHOR(S)
[editorial board member report]

I have been asked to assess the submitted Comment below, as well as another Comment by the same author for another paper by Lin et al.

I have concerns with this approach to communicating disagreements with published work. The Comments focus on differences of opinion regarding clinical feasibility and read like a referee report ought to. The Commenting author may not approve of the assumptions made or the specific approaches of Lin et al., but if there are no major errors, I don't see the need for publication of the Comment. I would suggest that the Commenter should write their own original article approaching the same topic, but it appears that they already have.

The purpose of the Comments seem to be to alert readers that the published work of Lin et al. used unrealistic assumptions, and therefore the results should not be trusted. However, in my opinion, Monte Carlo simulations of nanoparticle enhancement all suffer from some sort of unrealistic assumptions and still generally underestimate the biological effects of nanoparticles in radiation therapy anyways. Arguing over these minor points rather than addressing the bigger issues facing this therapy does not seem like the best use of time or resources.

I also wonder whether Lin et al. will be given an opportunity to respond to the criticisms in the Comments and what the mechanics of that will be. Ultimately, this could be an interesting discussion, but I don't think that it is appropriate to have that play out over months/years through this Comment mechanism. What I would like to see is for the Comments' author and Lin et al. to collaborate offline to produce a follow up article addressing the issues together.

Ultimately, the question I come back to is: "how will these Comments help patients with cancer?" I don't really see that they do, and therefore I don't see a need for publication.

Letter reference: DEC:RejBM:S

**Antwort: Your manuscript PMB-114487 - Decision on your manuscript**

Hans Rabus an pmb

27.01.2023 18

Details verberg

| Von | Hans Rabus/PTB |
| --- | --- |
| An | pmb@ioppublishing.org |

Dear Editor,

thank you for your email informing me of your decision on my submitted Comment.

Given that my motivation for writing the Comments is fully aligned with some of the considerations that make Reviewer 2 recommend rejection, I make another attempt of rebuttal.

Reviewer 2 is concerned that publishing the Comment could be the start of a futile debate that might be a waste of resources that would be of no use for patients with cancer.

The fear that a feud would be started is completely unsubstantiated. The referees' reactions to the Comment are in clear contrast to those of the two co-authors of the papers, who supervised the lead author when she was working at MGH before quitting scientific research. Harald Paganetti's replied that "Such discussions are important" and Jan Schuemann even offered me to become a co-author of the Comment to give it more weight. Although such an approach would have avoided any potential concerns about malevolent intentions on my side, most of the reviewers' arguments could then have been raised as well. A confirmation of the caveats raised by the Comment in a Reply by the authors would give them even more weight in my view. In any case, I am confident that the discussion will be brief, factual and very constructive.

The main motivation for me to write the Comments in the first place was that resources are wasted when other authors copy approaches and assumptions of papers like those of Lin et al. without being aware of the caveats. "The Comments read like a referee report should" because they are extracts of a detailed referee report I have recently written for another journal in the area of medical physics, where I reviewed a manuscript that was largely inspired by the papers of Lin. et al. 2014, 2015 in PMB and essentially copied their methodology and choice of parameters of the simulations.

I fully agree with the reviewer that many published Monte Carlo simulations of nanoparticle enhancement generally underestimate the biological effects nanoparticles may have in radiation therapy. At least, when gold nanoparticle concentrations are considered that have actually been reported so far in preclinical or radiobiological studies. (There are exceptions such as McMahon et al., Radiother Oncol 100 (2011) 412–416.)

I disagree that this failure of Monte Carlo simulations is due to unrealistic simplifications. From my experience as reviewer and supervisor of students, the vast majority of these studies suffer from either considering extreme cases where giant effects were found, such as with the papers of Lin et al., or methodological deficiencies that end up comparing apples with oranges. In consequence, conflicting conclusions can be found in literature on questions such as optimum nanoparticle size and colleagues trying to review the field (e.g., Moradi et al., RPC 180 (2021) 109294) have to limit their endeavors to simply reporting "A did M and obtained X while B did N and obtained Y".

For a patient suffering from cancer today it does not make a difference whether the Comments are published or not. But this "no immediate benefit" catch may apply to many papers published in PMB. Suppressing the Comments means continuing the waste of resources and efforts of early-stage researchers or other newcomers to the field who copy the methodology of published work not knowing that there are caveats and plausibility concerns. This will perpetuate the confusion in literature and hamper progress in our understanding of nanoparticle radiation effects and delay the full exploitation of this promising therapy.

Starting the discussion now gives us the possibility to change this situation, make the field focus on the real challenges and eventually make the introduction of nanoparticles in clinical practice happen earlier and have future patients suffering from cancer benefit from this modality. Suppressing the scientific discussion will block this way.

It is up to the Editorial board to decide which alternative PMB should support.
With kind regards

Hans Rabus

### A message about manuscripts PMB-114492 & PMB-114487
Physics in Medicine and Biology
An hans.rabus

10.02.2023 13:04
Details verbergen

Von "Physics in Medicine and Biology" <onbehalfof@manuscriptcentral.com>
An hans.rabus@ptb.de
Bitte Antwort an pmb@ioppublishing.org

Dear Dr Rabus,

Re: "Comment on "Comparing gold nano-particle enhanced radiotherapy with protons, megavoltage photons and kilovoltage photons: A Monte Carlo simulation" by Lin et al [Phys. Med. Biol. 59 (2014) 7675–7689]"

Manuscript reference: PMB-114492

and

"Comment on "Biological modeling of gold nanoparticle enhanced radiotherapy for proton therapy" by Lin et al. [Phys. Med. Biol. 60 (2015) 4149–4168]"

Manuscript reference: PMB-114487

Thank you for your latest message regarding the two Comment articles you submitted.

Two Editorial board members have independently recommended that we do not consider your Comment articles. After further consultation with the Editorial board I can confirm that we will not be able to reconsider that decision.